\documentclass[reprint,aps,prl,amsmath,amsfonts,amssymb]{revtex4-1}
\usepackage{natbib}
\usepackage{amsmath}
\usepackage{amsfonts}
\usepackage{amssymb}
\usepackage{wasysym}
\usepackage{mathrsfs}
\usepackage{graphicx}
\usepackage{times}
\DeclareMathOperator{\Tr}{\text{Tr}}
\renewcommand{\vec}[1]{\mathbf{#1}}
\newcommand{\im}[1]{\operatorname{Im}\left[#1\right]}
\newcommand{\re}[1]{\operatorname{Re}\left[#1\right]}
\newcommand{\tr}[1]{\operatorname{Tr}\left\{#1\right\}}

\newcommand{\asym}[1]{\operatorname{Asym}\left\{#1\right\}}

\renewcommand{\b}{\beta}

\newcommand{\add}[1]{\if\a\b{{\color{red} #1}}\else{#1}\fi}

\newcommand{\bracket}[1]{\langle #1 \rangle}
\newcommand{\ket}[1]{| #1 \rangle}
\newcommand{\bra}[1]{\langle #1 |}

\usepackage{lipsum}
\begin{document}
\title{$\mathbb{T}$-operator bounds on angle-integrated absorption and thermal radiation for arbitrary objects}
\author{Sean Molesky}
\author{Weiliang Jin}
\author{Prashanth S. Venkataram}
\author{Alejandro W. Rodriguez}
\affiliation{Department of Electrical Engineering, Princeton University, Princeton, New Jersey 08544, USA}
\begin{abstract}
\noindent
We derive fundamental per-channel bounds on angle-integrated absorption and thermal radiation
for arbitrary bodies---for any given material susceptibility and bounding region---that simultaneously encode both the per-volume limit on polarization set by passivity and geometric constraints on radiative efficiencies set by finite object sizes through the scattering $\mathbb{T}$-operator. 
We then analyze these bounds in two practical settings, comparing against prior limits as well as near optimal structures discovered through topology optimization. 
Principally, we show that the bounds properly capture the physically observed transition from the volume scaling of absorptivity seen in deeply subwavelength objects (nanoparticle radius or thin film thickness) to the area scaling of absorptivity seen in ray optics (blackbody limits).
\end{abstract}
\maketitle
\noindent
Motivated by the increasing control of light offered by micro and nanoscale structuring~\cite{koenderink2015nanophotonics,molesky2018inverse}, impetus to find bounds analogous to the blackbody limit for geometries that violate the assumptions of ray optics (nanoparticles~\cite{du2017near}, thin films~\cite{tong2015thin}, photonic crystals~\cite{zhu2015radiative,ilic2016tailoring}, etc.) has steadily grown over the past few decades. 
It is now well established that the absorption (radiative thermal emission) cross-sections of a compact object can be much greater than its geometric area~\cite{bohren2008absorption,tribelsky2011anomalous,ruan2011design,biehs2016revisiting,fernandez2018super,thompson2018hundred} (``super-Planckian'' emission), and that deeply subwavelength films can achieve near unity absorptivity via surface texturing~\cite{kats2016optical,dyachenko2016controlling}. 
Limits applicable to all length scales and materials could both provide insight into these representative phenomena and guide efforts in related application areas such as integrated and meta-optics~\cite{kruk2017functional,khorasaninejad2017metalenses,hampson20183d}, photovoltaics~\cite{yablonovitch1982statistical,atwater2010plasmonics,zhang2016near,jariwala2017van}, and photon sources~\cite{thompson2013coupling,galfsky2015active,somaschi2016near}. 

Development of bounds for arbitrary objects have primarily followed two overarching strategies: modal decompositions based on quasi-normal, singular value, Fourier and/or multipole expansions~\cite{mclean1996re,hamam2007coupled,kwon2009optimal,yu2010fundamental,yu2010grating,ruan2010superscattering,hugonin2015fundamental,jia2015theory,yang2017low}, relating absorption cross-section to the number of excitable optical modes (channels); or material/region bounds, utilizing energy~\cite{callahan2012solar,miller2016fundamental} and/or spectral sum rules~\cite{fuchs1976sum,miller2000communicating,miller2007fundamental,miller2014fundamental,shim2019fundamental} to constrain achievable properties. 
Separately, each of these approaches present challenges for photonic design. 
Modal decompositions incorporate the specific size and shape characteristics of a body through expansion coefficients, and hence, inherently, require some enumeration and characterization of the participating modes to determine the range of values these coefficients can take~\cite{hamam2007coupled,ruan2012temporal,alpeggiani2017quasinormal}. 
Although fundamental considerations (transparency, energy, size, etc.) can and have been used in this regard~\cite{miller2000communicating,sohl2007physical,yu2010fundamental,yang2017low}, such cut-offs have yet to tightly bound potential coefficient values for arbitrary compact geometries, particularly when applied to metallic nanoparticles and antennas~\cite{gustafsson2007physical,kwon2009optimal,miller2016fundamental}. 
Conversely, material bounds set by intrinsic dissipation naturally reproduce the volumetric scaling of absorptivity characteristic of deeply subwavelength objects (and are highly accurate for the special case of weak polarizability in this regime~\cite{miller2016fundamental}). 
However, because such approaches intrinsically suppose an optimally large response field existing at all points within an arbitrary object for any incident field, the same volumetric scaling persists for all length scales. 
Consequently, material bounds can rapidly become too loose beyond quasi-static settings, yielding unphysical divergences with both increasing object size and material response.

In this letter, we derive bounds on thermal radiation and absorption that combine these two approaches, linking the impact of material response with the influence of an object's geometry through the scattering $\mathbb{T}$-operator. 
This leads to a per-channel limit on integrated absorption capturing both extraction and radiative (scattering) loss processes through the singular values of the imaginary part of the vacuum Green function. 
The result is applicable to objects of any size, exhibiting a smooth transition in absorptivity from the volume scaling achievable in the quasi-static (deeply-subwavelength) regime to the area scaling limit of macroscopic ray optics. 
Further, the bounds always asymptotically approach the ray optics limit (when all characteristic lengths are large) and diverge sub-logarithmically (rather than linearly) with material quality for objects of finite extent, significantly reducing cross-section limits for typical optical media even when all characteristic lengths are small. 
Throughout, we compare the present results to prior bounds as well as structures discovered using topology optimization, realizing a variety of examples (metallic and dielectric) that nearly achieve the predicted limits.
\\ \\
\textbf{Derivation---}From the relations of scattering theory, both the power scattered from an incident field ($\left|\textbf{E}_{\text{inc}}\right>$), and the thermal radiation emitted at temperature $T$, can be expressed in terms of the scattering $\mathbb{T}$-operator of an object and the vacuum Green function $\mathbb{G}^\mathrm{vac}$~\cite{kruger2012trace} as
\begin{align}
  &P_{\text{sct}}\left(\omega\right) =\frac{k_{o}}{2Z} \left<\textbf{E}_{\text{inc}}\right|\mathbb{T}^{\dagger}\im{\mathbb{G}^{\mathrm{vac}}}\mathbb{T}\left|\textbf{E}_{\text{inc}}\right>
  \nonumber \\
  &=\frac{k_{o}}{2Z} \left<\textbf{E}_{\text{inc}}\right|\im{\mathbb{T}} - \mathbb{T}^{\dagger}\im{\mathbb{V}^{-1\dagger}}\mathbb{T}\left|\textbf{E}_{\text{inc}}\right>,
  \label{Psct}
 \end{align}
 and,
 \begin{align}
  &H\left(\omega,T\right) = \Pi\left(\omega,T\right)\Phi\left(\omega\right)
  \nonumber \\
  &\Phi = \frac{2}{\pi}\Tr\left[\im{\mathbb{G}^\mathrm{vac}} \left(\im{\mathbb{T}} - \mathbb{T}^{\dagger}\im{\mathbb{G}^\mathrm{vac}}\mathbb{T}\right)\right].
  \label{preAbs}
\end{align}
Here, $\omega$ is the angular frequency,  $k_{o} = 2\pi/\lambda$ is the wavenumber, $Z$ is the impedance of free-space,  $\Pi\left(\omega\right) = \hbar\omega /\left(\text{Exp}\left(\beta \hbar\omega\right) - 1\right)$ with $\left(\beta = 1/k_{\text{B}}T\right)$ is the Planck energy of a harmonic oscillator, $\text{Tr}\left[\ldots\right]$ denotes the trace, $\im{\mathbb{T}} = \left(\mathbb{T}-\mathbb{T}^{*}\right)/ 2i$, and, by Kirchhoff's law of thermal radiation, $\Phi$ is the object's angle integrated absorption~\cite{greffet1998field}. 
(A synopsis of scattering formalism, along with a derivation of \eqref{preAbs}, is provided in Supplemental Material~\cite{supplement}.) 
For a passive object, scattered power must be positive for any incident field. 
As such, \eqref{Psct} simultaneously dictates that all singular values of the $\mathbb{T}$-operator must be smaller than the \emph{material figure of merit} $\zeta$
\begin{equation}
  \left\lVert\mathbb{T}\right\rVert \leq \zeta = \frac{\left|\chi\left(\omega\right)\right|^2}{\text{Im}\left[\chi\left(\omega\right)\right]},
  \label{materialMax}
\end{equation}
which was similarly derived in Ref.~\cite{miller2016fundamental} for polarization fields, and that $\im{\mathbb{T}}$ is positive-definite.

As $\im{\mathbb{G}^\mathrm{vac}}$ is real-symmetric positive-definite, it can be expressed via a singular value decomposition as
\begin{equation} 
  \im{\mathbb{G}^\mathrm{vac}} = \sum_{i} \rho_{i}\left| \textbf{q}_{i} \rangle \langle \textbf{q}_{i} \right|,
  \label{GvacSVD}
\end{equation}
where, as supported by our later analysis, each $\rho_{i}$ (eigenvalue) can be equated to the outgoing radiative flux of the $i$th mode---the $i$th \emph{radiative efficacy} of the domain. 
(The set $\left\{\rho_{i}\right\}$ plays an analogous role to the coupling coefficients $\left\{g_{ij}\right\}$ used by D.A.B. Miller in setting limits on far-field optical communication~\cite{miller2000communicating}.) 
Consider \eqref{preAbs} using this expansion, $\Phi = \left(2/\pi\right)\sum_{i}\rho_{i}\im{\langle \textbf{q}_{i}\left|\mathbb{T}\right| \textbf{q}_{i}\rangle}-\rho_{i}^{2}\left|\langle \textbf{q}_{i}\left|\mathbb{T}\right| \textbf{q}_{i}\rangle\right|^{2}-$ $\left(2/\pi\right)\sum_{\left\{\left(i,j\right) |i\neq j\right\}}\rho_{i}\rho_{j}\left|\langle \textbf{q}_{i}\left|\mathbb{T}\right|\textbf{q}_{j}\rangle\right|^{2}.$ 
Now, take $\mathbb{T}_{\text{opt}}$ to be a general operator described by the properties $(\mathbb{T}_{\text{opt}})^{\text{T}} = \mathbb{T}_{\text{opt}}$ (reciprocity), $\left\lVert\mathbb{T}_{\text{opt}}\right\rVert \leq \zeta$ (passivity), and $\im{\mathbb{T}_{\text{opt}}}$ positive-definite (passivity), ignoring all other physical constraints that any true $\mathbb{T}$-operator must satisfy. 
In this context, two characteristics of any maxima of $\Phi$ are clear. 
First, as $\left(\forall ~i, j\right)~\rho_{i}\rho_{j}\left|\langle\textbf{q}_{i}\left|\mathbb{T}_{\text{opt}}\right|\textbf{q}_{j}\rangle\right|^{2}\geq 0,$ the appearance of any cross-terms ($\langle\textbf{q}_{i}\left|\mathbb{T}_{\text{opt}}\right|\textbf{q}_{j}\rangle$) will always decrease $\Phi$. 
Therefore, to maximize $\Phi$, a general operator $\mathbb{T}_{\text{opt}}$ must be diagonalized in the basis of $\im{\mathbb{G}^\mathrm{vac}}$, \eqref{GvacSVD}. 
Second, the complex phase of $\left<\textbf{q}_{i}\left|\mathbb{T}_{\text{opt}}\right|\textbf{q}_{i}\right>$ only influences the first (positive) piece of the sum, and so the value of $\Phi$ peaks when $\left(\forall i\right)~ \text{atan}\left(\im{\langle\textbf{q}_{i}\left|\mathbb{T}\right|\textbf{q}_{i}\rangle}/\re{\langle\textbf{q}_{i}\left|\mathbb{T}\right|\textbf{q}_{i}\rangle}\right)= \pi/2.$ 
Together, these two considerations show that achievable values of $\Phi$ are bounded by taking $\mathbb{T}_{\text{opt}}$ to be diagonalized by \eqref{GvacSVD} with purely imaginary eigenvalues: $\mathbb{T}_{\text{opt}} = \sum_{i} i\tau_{i}\left|\textbf{q}_{i}\rangle \langle \textbf{q}_{i}\right|$ with $\left(\forall i\right)~\tau_{i} \in [0,\zeta]$. 
As such,
\begin{equation}
  \Phi_{\text{opt}} =\frac{2}{\pi} \sum_{i}\tau_{i}\rho_{i} -\left(\tau_{i}\rho_{i}\right)^{2}
  \label{absorptionForm}
\end{equation}
and maximizing the contribution of each $\tau_{i}$ yields
\begin{align}
  \Phi_{\text{opt}} &=\frac{2}{\pi} \sum_{i}
    \begin{cases}
        1/4  & \left(\zeta\rho_{i} \geq 1/2\right)\\
        \zeta\rho_{i} - \left(\zeta\rho_{i}\right)^{2} &\text{else}.
    \end{cases}
  \label{absorptionBound}
\end{align}
That is, based on the criterion $\zeta \rho_{i} \geq 1/2$, each channel in \eqref{absorptionBound} produces either the Landauer limited contribution of $1/4$~\cite{molesky2019limits}, or the material limited $\zeta\rho_{i}-\left(\zeta\rho_{i}\right)^{2}$. 
\\ \\ 
\textbf{Interpretation---}In terms of the $\mathbb{T}$ operator, the total power extracted from any incident field $\left|\textbf{E}_{\text{inc}}\right>$ by an object is $P_{\text{ext}}
=k_{o}\left<\textbf{E}_{\text{inc}}\right|\im{\mathbb{T}}\left|\textbf{E}_{\text{inc}}\right>
/ \left(2Z\right)$. 
Comparing with \eqref{Psct} and \eqref{preAbs}, $\Phi$ thus amounts to the difference of the extracted ($\im{\mathbb{T}}$) and scattered ($\mathbb{T}^{\dagger}\im{\mathbb{G}^\mathrm{vac}}\mathbb{T}$) power for free-space states. 
The separation of these two forms persists throughout the derivation of the bounds, representing the linear and quadratic terms of \eqref{absorptionForm}. 
$\Phi_{\text{opt}}$ results from their connected physics.

In real space, $\tr{\im{\mathbb{G}^\mathrm{vac}}} = \sum_{i}\rho_{i}$ is the integral of the local density of free-space states over the domain of the object. 
Following \eqref{absorptionBound}, the total power that can be extracted by an object, the first term of \eqref{preAbs}, is thus bounded by its ability to interact with radiative modes, $\tr{\im{\mathbb{T}}\im{\mathbb{G}^\mathrm{vac}}}=\sum_{i}\tau_{i}\rho_{i}$, which is maximized (independently) under complete saturation of material response, $\left(\forall i\right)~\tau_{i}=\zeta$. 
Relatedly, this form is also the result of applying the per-volume (shape independent) optical response limit of Ref.~\cite{miller2016fundamental} to integrated absorption, and is similar to the light trapping bound of Ref.~\cite{callahan2012solar}. 
Due to these connections with prior work,
\begin{equation}
  \Phi_{\text{qs}}\left(\omega\right) = \sum_{i}\zeta\rho_{i} = 
        \zeta\int\limits_{V}d\textbf{r}\im{\mathbb{G}^\mathrm{vac}\left(\textbf{r},\textbf{r}\right)}
  \label{homBound}
\end{equation}
serves as a useful comparison for $\Phi_{\text{opt}}$, and is subsequently referred to as the \emph{quasi-static bound}. 
This name is chosen as \eqref{homBound} follows from the assumption that the interaction of the object with any incident field is identically material limited, which can occur in quasi-static settings. 
This does \emph{not} mean that $\Phi_{\text{qs}}$ is valid only under the quasi-static approximation. 
Like $\Phi_{\text{opt}}$, $\Phi_{\text{qs}}$ is a mathematical bound derived from Maxwell's equations, albeit for any selection of parameters $\Phi_{\text{opt}} \leq \Phi_{\text{qs}}.$ 
\begin{figure*}[t!]
  \centering
  \includegraphics{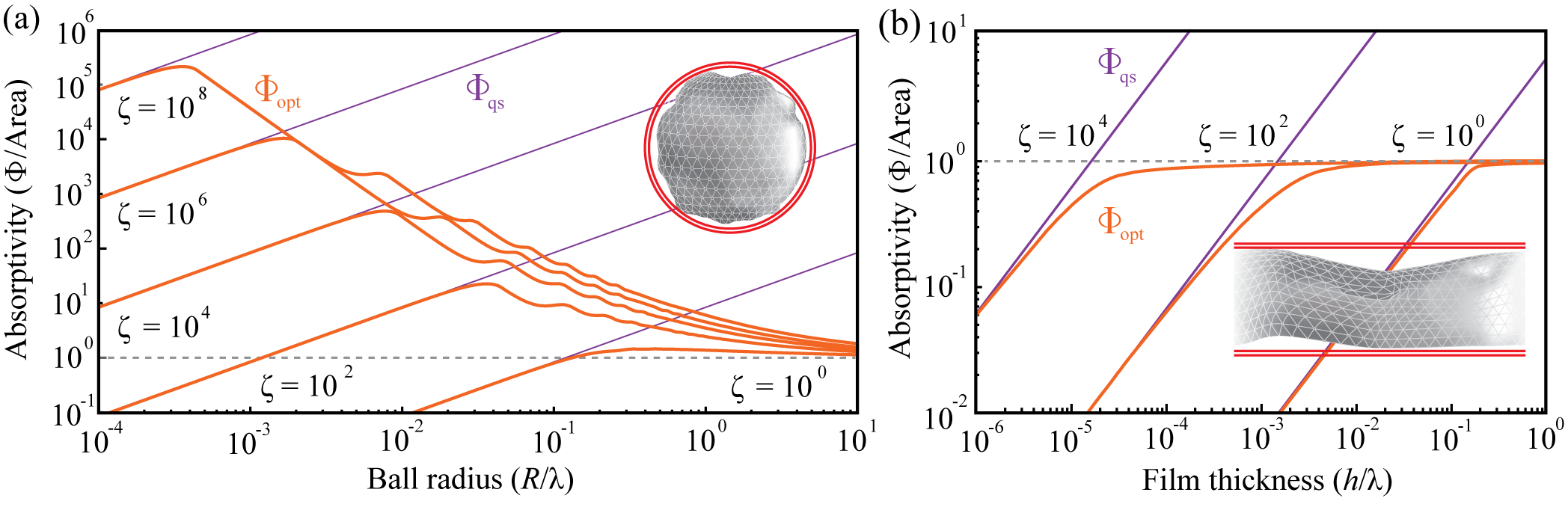}
  \vspace{-10 pt}
  \caption{\textbf{Bounds on angle-integrated absorption and thermal radiation for compact and extended bodies.} Absorptivity ($\Phi$ normalized by area $A$) bounds $\Phi_{\text{opt}}$ (orange lines) and $\Phi_{\text{qs}}$ (purple lines), for a range of $\zeta=|\chi|^2/\text{Im}\left[\chi\right]$ at a fixed wavelength $\lambda$. 
  These quantities are shown as a function of the wavelength normalized radius $R$ of an enclosing sphere (a), and thickness $h$ of a semi-infinite film (b). 
  Schematics of each setting are included as insets. 
  Even for small characteristic lengths ($\left\{R,h\right\}\leq 0.1\lambda$) $\Phi_{\text{opt}}$ is orders of magnitude smaller than $\Phi_{\text{qs}}$.}
  \vspace{0 pt}
\end{figure*}

In \eqref{absorptionForm} and \eqref{absorptionBound} this extracted power contribution is suppressed by scattering (radiative) losses, which are captured in the quadratic term in \eqref{preAbs} as the coupling of the polarization currents generated within an object back to free-space modes: originating through the operator $\mathbb{T}^{\dagger}\im{\mathbb{G}^\mathrm{vac}}\mathbb{T}$, each $\tau_{i}$ represents the ability of the object to convert a given field into a current, and each radiative efficacy $\rho_{i}$ the conversion of a current into outgoing radiative flux. Equivalently, the presence of strong polarization currents, necessary for strong per-volume absorption, leads to radiative loses, and these loses limit possible absorption. 
If $\zeta\rho_{i} > 1/2$, mirroring the observed dependencies of absorption ($\propto V$) and scattering ($\propto V^{2}$) seen in highly subwavelength metallic antennas~\cite{hamam2007coupled,ruan2010superscattering}, the growth of radiative losses with increasing $\tau_{i}$ can potentially
surpass the growth of the extracted power, inducing saturation. 
As both processes are rooted in the same conversion between radiative fields and polarization currents, this critical coupling occurs at the compelling value of $\tau_{i}\rho_{i} = 1/2$~\cite{pendry1983quantum,pendry1999radiative}, the probability of a maximally entropic Bernoulli process, resulting in the Landauer limit value of $\Phi_{\text{opt}}=1/4$. 
\\ \\ 
\textbf{Analysis---}The practical usefulness of \eqref{absorptionBound} stems from its favorable mathematical properties. 
Namely, \eqref{absorptionBound} monotonically increases with $\zeta$ or any $\rho_{i}$, and, as proved in Supplementary Material~\cite{supplement}, each $\rho_{i}$ increases if the object grows (domain monotonicity). 
This allows us to freely decouple any true object from an imagined encompassing region of space (bounding domain). 
A mismatch between the domain of the object and the domain of $\im{\mathbb{G}^\mathrm{vac}}$ must technically reduce $\left\lVert\mathbb{T}\right\rVert$ below $\zeta$, but without any modification \eqref{absorptionBound} remains an upper bound on $\Phi$. 
That is, the result of \eqref{absorptionBound} for any particular bounding domain is applicable to any object that can be enclosed (as well as any sub-domain).

The procedure for calculating $\Phi_{\text{opt}}$ is straightforward for any bounding geometry (e.g. wires, disks, spheres, extended films, stars, disconnected patches, etc.). 
Precisely, the set of singular values $\left\{\rho_{i}\right\}$ of the domain can always be computed by forming a real space matrix representation of $\im{\mathbb{G}^{\mathrm{vac}}}$, 
\begin{align}
  &\im{\mathbb{G}^{\mathrm{vac}}}\left(\textbf{r}\right) = 
  \frac{k_{o}^{3}}{4\pi r}\Bigg[\left(\text{sin}\left(r\right) + \frac{\text{cos}\left(r\right)}{r} - \frac{\text{sin}\left(r\right)}{r^{2}}\right)\overline{\text{I}} -\nonumber\\
  &\left(\text{sin}\left(r\right)+\frac{3~\text{cos}\left(r\right)}{r}-\frac{3~\text{sin}\left(r\right)}{r^{2}}\right)\hat{\textbf{r}}\otimes\hat{\textbf{r}}\Bigg],
\end{align}
with every $r$ multiplied by a hidden $k_{o}$, and then performing a singular value decomposition of the result~\cite{polimeridis2013directfn,polimeridis2015fluctuating}. 
Here, to facilitate further investigation, we will focus on the high symmetry case of a ball where semi-analytic evaluation is manageable (expressions for films, as well as minor additional details, are given in Supplemental Material~\cite{supplement}). 
Nevertheless, we stress that determining $\Phi_{\text{opt}}$ for domains lacking symmetry does not raise any meaningful computationally difficulties. 

For this geometry two types of singular values arise
\begin{align}
	\rho_{\ell}^{\left(1\right)} 
    &=\frac{\pi R^{2}}{4} \Bigg(\frac{\ell+1}{2\ell+1}\left(J_{\ell-\frac{1}{2}}^{2}\left(R\right) - J_{\ell+\frac{1}{2}}\left(R\right)J_{\ell-\frac{3}{2}}\left(R\right)\right) \nonumber \\ &\hspace{0.2in}+ \frac{\ell}{2\ell+1}\left(J_{\ell+\frac{3}{2}}^{2}\left(R\right) - J_{\ell+\frac{1}{2}}\left(R\right)J_{\ell+\frac{5}{2}}\left(R\right)\right)\Bigg),
    \nonumber\\
	\rho_{\ell}^{\left(2\right)} &= \frac{\pi R^{2}}{4}\left(J_{\ell+\frac{1}{2}}^{2}\left(R\right) - J_{\ell-\frac{1}{2}}\left(R\right) J_{\ell+\frac{3}{2}}\left(R\right)\right),
    \label{singVals}
\end{align}
where $J_{\ell}\left(-\right)$ is the $\ell$th Bessel function of the first kind with an additional factor of $2\pi$ included in its argument, each $\ell$ (spherical harmonic) index has a multiplicity of $\left(2\ell+1\right)$, and $R$ is the radius of the ball normalized by the wavelength. 
Using standard properties of Bessel functions, it can be shown that for values of $R \gg \ell$, each of these singular values tends to the asymptote  $2\pi^{2} R$, and that for any combination of arguments $\rho^{\left(1\right)}_{\ell} < \pi\left(\ell+1\right) \left(\pi R\right)^{2\ell+1}/\left(2 \Gamma^{2}\left(\ell+3/2\right)\right) + 2\pi \ell \left(\pi R\right)^{2\ell+5}/\left(\left(2\ell+5\right)\left(2\ell+3\right) \Gamma^{2}\left(\ell+5/2\right)\right)$ and $\rho^{\left(2\right)}_{\ell} < 2\pi \left(\pi R\right)^{2\ell+3}/\left(\left(2\ell+3\right)\Gamma^{2}\left(\ell+3/2\right)\right)$ (asymptotically approached for small values of $R$). 
These forms reveal two prescient general features. 
First, in the limit of small domains ($R \ll 1$), with ``small'' being determined by the value of $\zeta$, only the first singular value of the first type contributes, and this triply degenerate (dipole) mode is responsible for the initial volume scaling necessitated by the physical meaning of the bounds. 
Second, the radial growth of the singular values shows that the saturation condition (impact of radiative losses) plays a major role in limiting radiative thermal emission and integrated-absorption in wavelength scale volumes. 
(For $\zeta = 10^{6}$, Fig.~1 (a), radiative losses lead to order of magnitude deviations of $\Phi_{\text{opt}}$ from $\Phi_{\text{qs}}$ beyond $R\approx 0.003 \lambda$.) 
As visually confirmed by Fig.~1 panel (a), as the domain grows an increasing number of channels (multipoles) saturate causing ``steps'' to appear in $\Phi_{\text{opt}}$, and these steps lead to successively larger deviation with $\Phi_{\text{qs}}$ that ultimately regularize the initial volumetric scaling. 
Results for films, Fig.~1 (b), are qualitatively similar. 
However, since the domain is infinite, the steps associated with saturation are now blended into a continuum, and the large characteristic size limit is approached from below rather than above. 
From a practical perspective, the fact that $\Phi_{\text{opt}}$ can achieve near ideal absorptivity for very small film thickness and moderate values of $\zeta$ is quite remarkable, a finding that is tacitly supported by a number of recent studies in 2D materials and meta-surfaces~\cite{thongrattanasiri2012complete,akselrod2015large,kim2018electronically,nong2018perfect}. 
Crucially, in either case, for any value of $\zeta$, $\Phi_{\text{opt}}$ asymptotes to a geometric perfect absorber (the blackbody limit). 

The asymptotic behavior of the singular values also reveals general characteristics of the dependence of $\Phi_{\text{opt}}$ on the material figure of merit $\zeta$. 
Applying Sterling's approximation to the bounding expressions given above, for $\left(\ell\gg e\pi R\right)$ we have $\rho^{\left(2\right)}_{\ell} \approx \left(e\pi R/\ell\right)^{2\ell+1}/4$ and $\rho^{\left(1\right)}_{\ell} \approx \left(e\pi R/\ell\right)^{2\ell+3}/2$, to arbitrary accuracy as $\ell$ becomes large. 
Fix $R$, and suppose that $\zeta = \rho^{\left(2\right)}_{k}$ ($\rho^{\left(1\right)}_{k}$ is 
analogous). 
Using the fact that $e\pi R/\left(k+\ell\right) < e\pi R/k$ the remaining (unsaturated) linear contribution of $\Phi_{\text{opt}}$ is then bounded by $9\left(e\pi R\right)^{3}/\left(4\left(k^{2}-\left(e\pi R\right)^{2}\right)\right)$. 
Hence, as $\zeta$ saturates increasingly higher spherical harmonics, the contribution of the remaining unsaturated harmonics becomes increasingly small compared to the contribution of the newly saturated harmonic, $\approx\left(2k +1\right)/4$. 
But, saturation of the $\ell$th singular value (in the large $\ell$ limit) requires
\begin{equation}
  \text{ln}\left(\frac{\zeta}{2}\right) \geq \left(2\ell + 1\right)\text{ln}\left(\frac{\ell}{e \pi R}\right),
  \label{matScale}
\end{equation}
which has a sub-logarithmic dependence between $\ell$ and $\zeta$. 
Due to domain monotonicity, the above material scaling result for a ball is applicable to all compact (finite sized) objects. 

This bound on material quality scaling is well matched to the features of the $\Phi_{\text{opt}}$ curves in Fig.~1 panel (a). 
Once the radius has surpassed $\approx\lambda$, geometric increases in $\zeta$ ($\times 10^{2}$) produce relatively minute changes in the bounds. 
This behavior also appears for smaller radii at larger values of $\zeta$, but this range is not of great practical relevance since materials with $\zeta$ surpassing $\approx 10^{8}$ are quite rare. For instance, in the optical to infrared, $\omega \in (0.5\text{--}15)\mu\text{m}$, $\zeta\left(\omega\right)$ has a peak value of approximately $1.7\times 10^{3}$ for gold, $2.4\times 10^{3}$ for tungsten, $2.2\times 10^{4}$ for silicon carbide, $6.8\times 10^9$ for silicon, $3.3\times 10^{7}$ for gallium arsenide, and $5.9\times 10^{7}$ for gallium phosphide~\cite{palik1998handbook}. 
\begin{figure}[ht!]
  \centering
  \includegraphics{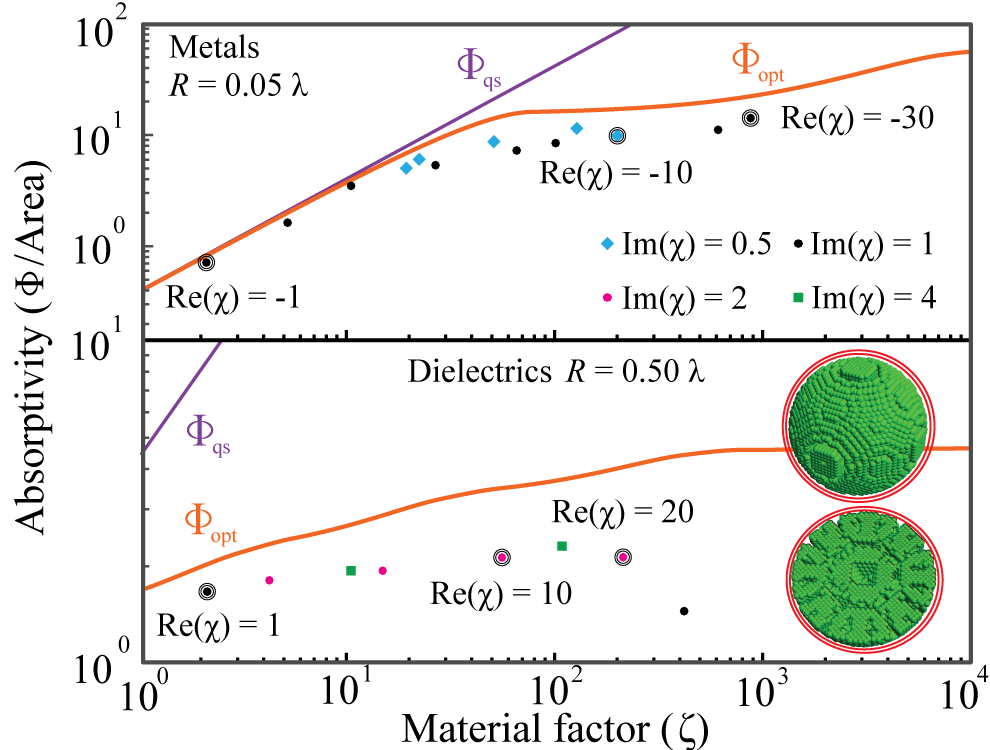}
  \vspace{-10 pt}
  \caption{\textbf{Comparison of bounds with geometries discovered by inverse design}. Absorptivity ($\Phi$ over area $A$) of structures discovered using gradient topology optimization for a variety of metallic (a) and dielectric (b) materials characterized by the material figure of merit $\zeta=|\chi|^2/\im\chi$. 
  (See text for more information.) 
  For comparison, the bounds $\Phi_{\text{opt}}$ \eqref{absorptionBound} and $\Phi_{\text{qs}}$ \eqref{homBound} are also depicted. 
  In (a), all structures are bound by a ball of radius $R=0.05\lambda$. 
  For panel (b), the confining domain is a ball of $R=0.5\lambda$. 
  The inset provides a visualization of the structure (exterior and planar cut) for the rightmost green square. 
  The observation that optimized structures come within factors of unity of $\Phi_{\text{opt}}$ provides case evidence of the tightness of \eqref{absorptionBound}.}
  \vspace{0 pt}
\end{figure}
\\ \\ 
\textbf{Optimizations---}Case evidence for the tightness of \eqref{absorptionBound} is presented in Fig.~2. 
Using a gradient topology optimization algorithm~\cite{jensen2011topology,molesky2018inverse}, see Supplemental Material for details~\cite{supplement}, structures nearly achieving $\Phi_{\text{opt}}$ have been discovered for two widely different domain sizes ($R = 0.05\lambda$ and $R = 0.5\lambda$) and a variety of metallic and dielectric susceptibilities. 
In Fig.~2, these media are grouped by imaginary susceptibility, corresponding to four different values of $\im{\chi}$, $\left\{0.5,1,2,4\right\}$, with the remaining variation in $\zeta$ occurring due to $\re{\chi}$. 
Explicit values of $\re{\chi}$ are given for circled points, providing a sense of the range considered. 
As was previously remarked by O. Miller et al.~\cite{miller2016fundamental}, $\Phi_{\text{qs}}$ is attained for a plane wave polarized along the axis of an ellipsoidal metallic nanoparticle, given a properly chosen aspect ratio. 
For small values of $\zeta$ this ratio is near unity and resonant metallic structures ($\re\chi \approx -3$) matching both bounds are easily discovered. 
As $\zeta$ moves to moderate values, the aspect ratio required for an ellipsoidal particle to match $\Phi_{\text{qs}}$ becomes increasingly extreme. 
Due to our chosen spherical boundary, discovered structures begin to deviate considerably from $\Phi_{\text{qs}}$, but continue to come within a factor of $2$ of $\Phi_{\text{opt}}$ up to $\zeta =
10^{3}$. 
Past this point, numerical issues impede our present algorithms and it remains to be seen how much of the roughly order of magnitude headroom allowed by $\Phi_{\text{opt}}$ is accessible.

Results for the larger domain, Fig.~2 (b), show similarly good agreement. 
An example structure is depicted in the right inset (full view and planar cut), corresponding to the rightmost green square in the plot. 
Comparing with the assumptions made in deriving \eqref{absorptionBound}, the $\mathbb{T}$ operator
for this structure ($\chi = 20 +4i$, $\Phi = 0.60~\Phi_{\text{opt}}$) is indeed found to be nearly diagonal in the basis of $\im{\mathbb{G}^\text{vac}}$ and have almost completely imaginary eigenvalues (for supporting data see Supplemental Material~\cite{supplement}). 
\\ \\
\textbf{Remarks}---There are a few points that should be considered when using \eqref{absorptionBound}, or comparing to prior literature. 
First, $\Phi_{\text{opt}}$ is a bound on thermal emission and integrated absorption for a given domain and $\zeta$ factor. 
By choosing different geometries and material parameters, \eqref{absorptionBound} can be applied to any desired context, but the confining volume is an essential feature. 
Second, there is no universal guarantee of tightness. 
Beyond the demonstrated agreement of the bounds with known quasi-static and ray optics asymptotics, the only a priori guarantee is domain monotonicity; there are likely volumes and material parameters where the value of $\Phi_{\text{opt}}$ will be larger than the true $\Phi$ of any practical structure. 
Next, while we have only considered single wavelengths, there is no reason the bounds can not be applied to finite frequency ranges. 
The derivation of $\Phi_{\text{opt}}$ presented above does not incorporate any spectral sum rules (derived from causality), such as the fact that $\mathbb{T}_{\text{opt}}$ should obey Kramers-Kronig dispersion relations, but for resonant absorption or thermal emission simply multiplying the bound by the width of the resonance should not produce a substantially looser bound than $\Phi_{\text{opt}}$ at the peak wavelength. 
(As an expedient, taking $\Phi_{\text{opt}}$ to be the peak value of a Lorentzian function of width $\Delta\omega = \omega~\mathrm{Im}\chi / |\chi|$ is likely a fair approximation.) 
Finally, as suggested in the introduction, $\Phi_{\text{opt}}$ can be interpreted as the extension of prior multipole analysis~\cite{mclean1996re,hamam2007coupled,kwon2009optimal,yu2010fundamental,yu2010grating,ruan2010superscattering,hugonin2015fundamental,jia2015theory,yang2017low}, or communication limits~\cite{miller2000communicating,miller2007fundamental}, to general domains with the crucial addition that an upper bound is set on the number modes which may contribute through the pseudo-rank of the imaginary part of the vacuum Green function ($\im{\mathbb{G}^\mathrm{vac}}$) and the material figure of merit ($\zeta$) \eqref{materialMax}. 
We foresee this rank revealing capability potentially providing a number of benefits for future practical design and optimization. 
We also note that much of what has been developed in this manuscript is applicable not only to generalized electromagnetic scattering (for incident planewaves or dipolar emitters with applications to solar cells, light-emitting diodes, and single-photon emitters), but also to quantum mechanics, acoustics, and other wave physics. 
\section{Acknowledgments}
\noindent
This work was supported by the National Science Foundation under Grants No. DMR-1454836, DMR 1420541, DGE 1148900, the Cornell Center for Materials Research MRSEC (award no. DMR1719875), the Defense Advanced Research Projects Agency (DARPA) under agreement HR00111820046, and the National Science and Engineering Research Council of Canada under PDF-502958-2017. 
The views, opinions and/or findings expressed herein are those of the authors and should not be interpreted as representing the official views or policies of any institution. 
We thank Jason Necaise for performing instructive calculations of the bounds in cylindrical coordinates. 
\section{Supplementary Information} 
\noindent
\textbf{$\mathbb{T}$-operator definition---}Following the Lippmann-Schwinger approach to scattering~\cite{lippmann1950variational}, the $\mathbb{T}$-operator is formally defined through the self-consistent equation
\begin{equation}
  \textbf{E}= \textbf{E}_\text{inc} + \mathbb{G}^\mathrm{vac}\mathbb{V}\textbf{E},
  \label{Lippmann-Schwinger}
\end{equation}
as
\begin{align}
  \mathbb{T}^{-1}=\mathbb{V}^{-1} -\mathbb{G}^\mathrm{vac},
        \qquad \textbf{J} =
        \mathbb{T}\mathbb{V}^{-1}\textbf{J}_\text{inc},
\end{align}
with the ``inc'' subscript denoting an initial (freely incident)
current (field), $\mathbb{G}^\mathrm{vac}$ the vacuum Green function,
and $\mathbb{V}$ a generalized electromagnetic susceptibility. A lack
of any subscript indicates total field quantities.
\\ \\
\textbf{Heat transfer to thermal emission---}In two recent (related) articles~\cite{JinPRB2019,molesky2019limits} we have established that the heat transfer between any two bodies, labeled $a$ and $b$, can be written in terms of the scattering $\mathbb{T}$-operator as 
\begin{align}
  \Phi = \frac{2}{\pi}\Tr\Big\{&\left(\left(\mathbb{M}_{b}\mathbb{T}_{b}\right)^{\dagger}\im{\mathbb{V}_{b}^{-1*}}\left(\mathbb{M}_{b}\mathbb{T}_{b}\right)\right) \nonumber \\
  &\left(\left(\mathbb{G}_{ba}^{\mathrm{vac}}\mathbb{T}_{a}\right)\im{\mathbb{V}_{a}^{-1*}}\left(\mathbb{G}_{ba}^{\mathrm{vac}}\mathbb{T}_{a}\right)^{\dagger}\right) \Big\}
  \label{heatT}
\end{align}
where $\mathbb{M}_{a}$ and $\mathbb{M}_{b}$ are mutual scattering operators defined as 
\begin{align}
  &\mathbb{M}_{a} = \left(\mathbb{I}_{a} - \mathbb{T}_{a}\mathbb{G}_{ab}^{\mathrm{vac}}\mathbb{T}_{b}\mathbb{G}_{ba}^{\mathrm{vac}}\right)^{-1}\nonumber \\
  &\mathbb{M}_{b} = \left(\mathbb{I}_{b} - \mathbb{T}_{b}\mathbb{G}_{ba}^{\mathrm{vac}}\mathbb{T}_{a}\mathbb{G}_{ab}^{\mathrm{vac}}\right)^{-1}.
\end{align}
To produce an expression for thermal emission, we will evaluate \eqref{heatT} in the limit that one of the objects, here chosen to be $b$, tends to a blackbody. Abstractly, a blackbody is an
encompassing, infinitely large, region capable of perfectly absorbing
any incident field: physically, a spherical
material shell of inner radius $r_{b}$ and outer radius $R_{b}$, in the simultaneous limit $r_{b}\rightarrow \infty$ and $\text{Im} ~\chi_{b}\rightarrow 0$ with $(R_{b} - r_{b}) \rightarrow\infty$ and $\text{Im} ~\chi_{b} \left(R_{b} - r_{b}\right)\rightarrow 1$. We begin by breaking \eqref{heatT} into
\begin{align}
  &\mathbb{O}_{b} = \mathbb{M}_{b}\mathbb{W}_{b}\im{\mathbb{V}_{b}}\mathbb{W}_{b}^{\dagger}\mathbb{M}_{b}^{\dagger}\nonumber \\
  &\mathbb{O}_{a} = \mathbb{G}_{ba}^{\mathrm{vac}} \left(\im{\mathbb{T}_{a}} - \mathbb{T}_{a}^{*}~\im{\mathbb{G}_{a}^{\mathrm{vac}}}\mathbb{T}_{a}\right)\mathbb{G}_{ab}^{^{\mathrm{vac}}*},
\end{align}
where $\mathbb{W}_{b}^{-1} = \mathbb{I}_{b} - \mathbb{V}_{b}\mathbb{G}_{b}$ is the current dressing operator (producing a total current from an initial current).
As $b$ tends to the blackbody limit, the above definition shows that $\mathbb{W}_{b}\rightarrow \mathbb{I}_{b}$, i.e. that the total electric current density tends to the free current density. Hence, $\mathbb{T}_{b} = \mathbb{V}_{b}\mathbb{W}_{b}$ tends to zero, and the mutual scattering operators $\mathbb{M}_{a}$ and $\mathbb{M}_{b}$ become $\mathbb{I}_{a}$ and $\mathbb{I}_{b}$ respectively. Therefore, 
\begin{equation}
  \mathbb{O}_{b}\rightarrow \asym{\mathbb{V}_{b}}
  \label{bbLimit}
\end{equation}
with $\im{\mathbb{V}_{b}}\rightarrow 0$ and the volume of $b$ becoming arbitrarily large. 

Now, for any collection of objects, the fluctuation dissipation theorem~\cite{eckhardt1984kirchhoff} states that 
\begin{align}
  &\iint\limits_{V_{b}~V_{c}} \mathbb{G}^{*}\left(\textbf{r}_{a},\textbf{r}_{b}\right)\im{\mathbb{V}}\left(\textbf{r}_{b},\textbf{r}_{c}\right)\mathbb{G}\left(\textbf{r}_{c},\textbf{r}_{d}\right) =\im{\mathbb{G}}\left(\textbf{r}_{a},\textbf{r}_{d}\right),
\end{align}
which in our notation becomes 
\begin{equation}
  \mathbb{G}^{*}~\im{\mathbb{V}}\mathbb{G} = \im{\mathbb{G}}.
\end{equation}
To apply this result to \eqref{heatT}, we imagine a fictitious addition to body $b$ that 
perfectly overlaps with $a$. 
Since $\im{\chi_{b}}\rightarrow 0$ and body $a$ has a finite extent, this addition will have no material effect on the value produced by \eqref{heatT}. However, its inclusion shows that consistent application of the blackbody limit must result in
\begin{equation}
  \lim_{\mathbb{V}_{b}\rightarrow 0}\mathbb{G}_{ab}^{*{\mathrm{vac}}}\im{\mathbb{V}_{b}}\mathbb{G}_{ba}^{\mathrm{vac}} = \im{\mathbb{G}_{a}^{\mathrm{vac}}}.
\end{equation}
Applying this result to \eqref{heatT}, using the simplification of \eqref{bbLimit}, we find that 
\begin{align}
  &\Phi = \frac{2}{\pi}\Tr\Bigg\{ \mathbb{G}_{ab}^{\mathrm{vac}*}\im{\mathbb{V}_{b}} \mathbb{G}_{ba}^{\mathrm{vac}} \left(\im{\mathbb{T}_{a}} - \mathbb{T}_{a}^{*}~\im{\mathbb{G}_{a}^{\mathrm{vac}}}\mathbb{T}_{a}\right)\Bigg\} \nonumber \\
  &= \frac{2}{\pi}\Bigg(\Tr\Bigg\{ \im{\mathbb{G}_{a}^{\mathrm{vac}}}  \left(\im{\mathbb{T}_{a}} - \mathbb{T}_{a}^{*}~\im{\mathbb{G}_{a}^{\mathrm{vac}}}\mathbb{T}_{a}\right)\Bigg\} -\lim_{\mathbb{V}_{a}\rightarrow 0}\nonumber \\
  &\tr{\mathbb{G}_{a}^{\mathrm{vac}*}\im{\mathbb{V}_{a}} \mathbb{G}_{a}^{\mathrm{vac}} \left(\im{\mathbb{T}_{a}} - \mathbb{T}_{a}^{*}~\im{\mathbb{G}_{a}^{\mathrm{vac}}}\mathbb{T}_{a}\right)} \Bigg)\nonumber \\
  &=\frac{2}{\pi}\Tr\Bigg\{ \im{\mathbb{G}_{a}^{\mathrm{vac}}}  \left(\im{\mathbb{T}_{a}} - \mathbb{T}_{a}^{*}~\im{\mathbb{G}_{a}^{\mathrm{vac}}}\mathbb{T}_{a}\right)\Bigg\},
  \label{heatEmission}
\end{align}
where in the middle expression $\mathbb{V}_{a}$ has been used for the blackbody limit inside body $a$. 
\\ \\
\textbf{Domain monotonicity--} Let $U$ be a subdomain of $D$, and  
\begin{align}
  &\im{\mathbb{G}^\mathrm{vac}_{_U}} = \sum_{i}u_{i}\left|\textbf{u}_{i}\rangle\langle\textbf{u}_{i}\right|\nonumber \\
  &\im{\mathbb{G}^\mathrm{vac}_{_D}} = \sum_{i}d_{i}\left|\textbf{d}_{i}\rangle\langle\textbf{d}_{i}\right|
\end{align} 
be the vacuum Green functions for each of these volumes. Assume that these sums are finite. The first singular vector, corresponding to the largest singular value, is equivalent to the complex vector field $\textbf{f}\left(\textbf{r}\right)$ maximizing the integral 
\begin{equation}
  \iint\limits_{V}d\textbf{r}d\textbf{r}'~\textbf{f}^{*}\left(\textbf{r}\right)\im{\mathbb{G}^\mathrm{vac}}\left(\textbf{r},\textbf{r}'\right)\textbf{f}\left(\textbf{r}'\right)
  \label{sing1Int}
\end{equation}
subject to the constraint,
$\int\limits_{V}d\textbf{r}~\textbf{f}^{*}\left(\textbf{r}\right)\textbf{f}\left(\textbf{r}\right)
= 1$. It is clear that moving to a larger volume is always favourable in
this context. In the worst case $\textbf{f}\left(\textbf{r}\right)$
simply does not change. Therefore, the largest singular value of
$\im{\mathbb{G}^\mathrm{vac}}$ is domain monotonic. Now, suppose
that all singular values up to $i=N$ have been shown to be domain
monotonic, so that $\left(\forall i \leq N\right) u_{i}\leq d_{i}$, and
consider $i = N+1$. Recall that the $N+1$ singular vector
$|\textbf{d}_{N+1}\rangle$ is defined by the property of maximizing
$\langle\textbf{d}_{N+1}\left|\im{\mathbb{G}^\mathrm{vac}}\right|\textbf{d}_{N+1}\rangle$
subject to the constraints $\left(\forall i \leq N\right) \langle
\textbf{d}_{i}|\textbf{d}_{N+1}\rangle = 0$ and $\langle
\textbf{d}_{N+1}|\textbf{d}_{N+1}\rangle = 1$. Take
$\pi_{_U}|\textbf{d}_{i}\rangle$ to be the projection of $|\textbf{d}_{i}\rangle$ onto the subdomain $U$. If $\sum_{i=1}^{N}\pi_{_U}|\textbf{d}_{i}\rangle$ spans
$\sum_{i=1}^{N}|\textbf{u}_{i}\rangle$, then we essentially return to
the case of the first singular vector. $|\textbf{u}_{N+1}\rangle$ is
orthogonal to all $\pi_{_U}|\textbf{d}_{i}\rangle \ni i\leq N$, and
zero outside $U$. Hence, it is orthogonal to all
$|\textbf{d}_{i}\rangle \ni i\leq N$. As $\langle
\textbf{u}_{N+1}\left|\im{\mathbb{G}^\mathrm{vac}_{_D}}\right|\textbf{u}_{N+1}\rangle
= \langle
\textbf{u}_{N+1}\left|\im{\mathbb{G}^\mathrm{vac}_{_U}}\right|\textbf{u}_{N+1}\rangle$,
with the added freedom of the additional volume we must have
$d_{N+1}\geq u_{N+1}$. If $\sum_{i=1}^{N}\pi_{_U}|\textbf{d}_{i}\rangle$ does not span
$\sum_{i=1}^{N}|\textbf{u}_{i}\rangle$, then some elements of
$\left\{|\textbf{u}_{i}\rangle \ni i\leq N\right\}$ are orthogonal to
all $\pi_{_U}|\textbf{d}_{i}\rangle$. Select one such vector and
denote it as $|\textbf{u}_{\alpha}\rangle$. Since $u_{\alpha}\geq
u_{N+1}$, by the preceding argument $d_{N+1}\geq u_{N+1}$.
\\ \\
\textbf{Singular values for balls---}To derive the singular values of
$\im{\mathbb{G}^\mathrm{vac}}$ for a ball we have followed the formulation in
Tsang et al.~\cite{tsang2004scattering}. In this work, it is shown
that
\begin{multline}
  \im{\mathbb{G}^\mathrm{vac}}\left(\textbf{r}_{a},\textbf{r}_{a}'\right) = k_{o}^{3}\sum\limits_{\ell=1}^{\infty}\sum\limits_{m=-l}^{\ell}\left(-1\right)^{m} \\
  \sum_{j=1}^{2} \textbf{rS}_{\ell,m}^{\left(j\right)}\left(r,\theta,\phi\right)\otimes\textbf{rS}_{\ell,-m}^{\left(j\right)}\left(r',\theta',\phi'\right)
  \label{GreenBall}
\end{multline}
where
\begin{align}
  &\textbf{rS}^{\left(1\right)}_{\ell m}\left(r,\theta,\phi\right) = \sqrt{\frac{2\ell +1}{4\pi \ell\left(\ell +1\right)}}~j_{\ell}\left(r\right)\textbf{V}^{\left(3\right)}_{\ell m}\left(\theta,\phi\right)\nonumber \\
  &\textbf{rS}^{\left(2\right)}_{\ell m}\left(r,\theta,\phi\right) = \sqrt{\frac{2\ell +1}{4\pi \ell\left(\ell +1\right)}}\Big(\frac{\ell\left(\ell +1\right)}{r}j_{\ell}\left(r\right)\textbf{V}^{\left(1\right)}_{\ell m}\left(\theta,\phi\right)\nonumber\\&+\frac{1}{r}\frac{d\left(r~j_{\ell}\left(r\right)\right)}{dr}\textbf{V}^{\left(2\right)}_{\ell m}\left(\theta,\phi\right)\Big),
\end{align}
and $j_{\ell}\left(R\right)$ is the $\ell$th spherical with additional factor of $2\pi$ included in its argument. In these definition, $\textbf{V}_{\ell m}^{\left(\alpha\right)}$ are the vector spherical harmonics, obeying the orthogonality conditions
\begin{multline}
  \int d\Omega ~\textbf{V}^{\left(\alpha\right)}_{\ell m}\left(\theta,\phi\right)\textbf{V}^{\left(\beta\right)}_{\ell 'm'}\left(\theta,\phi\right) \\ = \delta_{\alpha,\beta}\delta_{m,m'}\delta_{\ell ,\ell'}
  \begin{cases} 
    \left(-1\right)^{m}\frac{4\pi}{2\ell+1}, &\alpha = 1\\
      \left(-1\right)^{m}\frac{4\pi \ell \left(\ell+1\right)}{2\ell+1}, &\alpha = \left\{2,3\right\}
  \end{cases}
\end{multline}
(As the spherical harmonics are unaffected by projection into a
connected volume, the program given below is also valid for shells and other homeomorphic domains.)
Comparing \eqref{GreenBall} with a standard singular value
decomposition, we equate the value of the inner product of the vector
pairs forming the above outer products with the singular values of
$\im{\mathbb{G}^\mathrm{vac}}$. Performing the angular integrals over the
vector spherical harmonics, results in two types of singular values quoted in the main text
\begin{align}
  \rho_{\ell m}^{\left(1\right)} &= \int\limits_{0}^{R}dr~r^{2}~\left(\frac{\ell +1}{2l+1}j_{\ell -1}^{2}\left(r\right)+\frac{\ell }{2l+1}j_{\ell +1}^{2}\left(r\right)\right) \nonumber \\
    &=\frac{\pi R^{2}}{4} \Bigg(\frac{\ell +1}{2l+1}\left(J_{\ell -\frac{1}{2}}^{2}\left(R\right) - J_{\ell +\frac{1}{2}}\left(R\right)J_{\ell -\frac{3}{2}}\left(R\right)\right) \nonumber \\ &\hspace{0.2in}+ \frac{\ell }{2l+1}\left(J_{\ell +\frac{1}{2}}^{2}\left(R\right) - J_{\ell +\frac{5}{2}}\left(R\right)J_{\ell +\frac{3}{2}}\left(R\right)\right)\Bigg)\nonumber\\
  \rho_{\ell m}^{\left(2\right)} 
  &= \int\limits_{0}^{R}dr~r^{2}~j_{\ell }^{2}\left(r\right)
  \nonumber\\  
  &= \frac{\pi R^{2}}{4}\left(J_{\ell +\frac{1}{2}}^{2}\left(R\right) - J_{\ell -\frac{1}{2}}\left(R\right) J_{\ell +\frac{3}{2}}\left(R\right)\right).
\end{align}
\\ \\
\textbf{Singular values for films---}Calculation of the singular
values for films of thickness $h$ follows a similar procedure to that
of a ball. Using results from Tsang et al.~\cite{tsang2004scattering}
and Kr{\"u}ger et al.~\cite{kruger2012trace}, the imaginary part of
the Green function can generally be decomposed in terms of
``regularized'' spectral basis as $\im{\mathbb{G}^\mathrm{vac}} =
\sum_{j \in \mathrm{prop}} \ket{\vec{E}^{\mathrm{reg}}_{j}}
\bra{\vec{E}^{\mathrm{reg}}_{j}}$. The ``propagating'' basis functions
of this set are orthogonal to one other (though not necessarily
self-normalizable), and form a complete set both at the origin and at
infinity. In particular, considering the in-plane wavevector $\vec{k}
= k_{x} \vec{e}_{x} + k_{y} \vec{e}_{y} \in \mathbb{R}^{2}$, the basis
functions can be written as plane waves
$\vec{E}^{\mathrm{reg}}_{s,p}(\vec{k},\vec{r})$, with $s \in \{-1,
1\}$ denoting the parity and $p \in \{M, N\}$ the polarization. The
vacuum Green function in this basis then takes on the form
\begin{align}
  &\im{\mathbb{G}^\mathrm{vac}}(\vec{r},\vec{r}') = \nonumber \\
  &\sum_{p,s} \int_{|\vec{k}| \leq k_0} \vec{E}^{\mathrm{reg}}_{s,p}(\vec{k},\vec{r})\otimes
  \vec{E}^{\mathrm{reg}*}_{s,p}(\vec{k},\vec{r}')~\frac{d^{2} k}{(2\pi)^{2}},
\end{align}
with singular vector functions given by
\begin{align}
   \vec{E}^{\mathrm{reg}}_{+,M}(\vec{k}, \vec{r}) &=
   \frac{ik_{o}~e^{i\left(k_{x} x + k_{y} y\right)}}{\sqrt{2k_{z}} |\vec{k}|} (k_{y} \vec{e}_{x} - k_{x} \vec{e}_{y})\cos(k_{z} z) \nonumber \\ 
   \vec{E}^{\mathrm{reg}}_{-,M}(\vec{k}, \vec{r}) &=
   \frac{-ik_{o}~e^{i\left(k_{x} x + k_{y} y\right)}}{\sqrt{2k_{z}} |\vec{k}|} (k_{y} \vec{e}_{x} - k_{x} \vec{e}_{y})\sin(k_{z} z) \nonumber \\ 
   \vec{E}^{\mathrm{reg}}_{+,N}(\vec{k}, \vec{r}) &=
   \frac{e^{i(k_{x} x + k_{y} y)}}{\sqrt{2k_{z}}|\vec{k}|}\Big(|\vec{k}|^{2} \cos(k_{z} z) \vec{e}_{z} \nonumber \\ 
   &-i\vec{k}k_{z}\sin(k_{z} z)\Big) \nonumber \\ 
   \vec{E}^{\mathrm{reg}}_{-,N}(\vec{k}, \vec{r}) &=
   \frac{e^{i(k_{x} x + k_{y} y)}}{\sqrt{2k_{z}}|\vec{k}|}\Big(i\vec{k}k_{z}\cos(k_{z} z) \nonumber \\
   &-|\vec{k}|^{2} \sin(k_{z} z) \vec{e}_{z}\Big) 
\end{align}
in position space, where $k_{z} = \sqrt{k_0^2 - |\vec{k}|^{2}}$ is real and nonnegative by virtue of the restriction to propagating waves. The
inner products may then all be written as
\begin{equation}
  \bracket{\vec{E}^{\mathrm{reg}}_{s',p'}(\vec{k}'),
    \vec{E}^{\mathrm{reg}}_{s,p}(\vec{k})} = \rho_{s,p}(\vec{k})
  (2\pi)^{2} \delta^{2} (\vec{k} - \vec{k}') \delta_{s,s'}
  \delta_{p,p'}
\end{equation}
due to the orthogonality of these basis functions. This immediately
yields the desired singular values corresponding to each $\vec{k}$:
\begin{align}
  \rho_{\pm,M}(\vec{k}) &= \frac{k_{o}^{2} h}{4} \left(1\pm
  \frac{\sin(k_{z} h)}{k_{z}h}\right) \nonumber\\
  \rho_{\pm,N}(\vec{k}) &= \frac{k_{o}^{2}h}{4k_{z}} \left(1 \pm  \frac{\sin(k_{z} h)}{k_{z}h} \right)\mp\frac{\sin\left(k_{z}h\right)}{2}.
\end{align}
\noindent
\textbf{Inverse design---}To explore the largest possible design
space, the optimizations shown in the text result from the ``topology''
(density) approach~\cite{molesky2018inverse}, in which each pixel
(permittivity value) within the chosen bounding domain is considered
as an independent design parameter. The key to the tractability of
such large-scale optimizations is the use of gradient-based
optimization algorithms (the method of moving asymptotes is employed in our algorithm~\cite{svanberg2002class}). To make use of these approaches,
each pixel is initially treated as continuous. That is, at each position in the 
domain, we begin by assign a parameter $\lambda_{i}\in\left[0,1\right]$ and state that 
the susceptibility of that pixel is 
\begin{equation}
  \chi_{i} = \lambda_{i}\chi_{\text{mat}},
  \label{linChi}
\end{equation}
where $\chi_{\text{mat}}$ is the material susceptibility. An initial 
optimization is then carried out using NLOPT~\cite{johnson2014nlopt}, 
producing a ``gray'' structure in which many $\lambda_{i}$ take on 
intermediate ($\neq \left\{0,1\right\}$) values. In subsequent
optimizations (for the same structure) these values are then binarized 
(i.e. forcing $\lambda_{i} = 0$ or $\lambda_{i} = 1$) by enclosing $\lambda_{i}$
in function, $\chi_{i} = f_{n}\left(\lambda_{i}\right)\chi_{\text{mat}}$,
which is slowly changed from the linear relation given above to a smooth
approximation of the step function~\cite{jensen2011topology}.

The primary challenge of this type of inverse design approach for the problem we have 
considered lies in the cost of computing $\Phi$ for inhomogeneous medium, which typically
needs to be evaluated thousands of times in a single optimization iteration. 
To surmount this difficulty, we have exploited 
our previously discussed fluctuating volume current formulation~\cite{polimeridis2015fluctuating,jin2016temperature}. This framework allows for two major simplifications.
First, it removes the necessity of simulating space outside the bounding domain, 
as would be necessary to accurately calculate emitted thermal radiation or angle-integrated 
absorption using a finite difference approach. Second, 
it allows the central matrix-vector multiplication that is used by 
the iterative inversion solver~\cite{polimeridis2015fluctuating}, 
$\mathbb{G}^\mathrm{vac} U_{i}$ in \eqref{invCalc}, to be computed via fast-Fourier transforms. 
To further reduce computational cost, we also use the fact that 
$\im{\mathbb{G}^\mathrm{vac}}$ has low pseudo-rank. (In a similar spirit to 
the derivation of the optimal $\mathbb{T}$-operator given the main text,
we use foreknowledge of the number of modes that will possibly contribute to select an appropriate
algorithm and pre-allocate computational resources.)
This allows us to formulate the scattering inversion problem, \eqref{invCalc},
directly in terms of a singular value decomposition (SVD) of
$\im{\mathbb{G}^\mathrm{vac}}=Q\Sigma Q^{\dagger}$, which can be
approximated to any accuracy with efficient randomized
methods~\cite{hochman2014reduced}. (For a full discussion of this procedure see Refs.~\cite{halko2011finding,martinsson2011randomized,polimeridis2015fluctuating}.) Specifically, starting from an equivalent trace form of heat transfer to the one given in the main text, the singular value decomposition of 
$\im{\mathbb{G}^\mathrm{vac}}$ allows us to recast $\Phi$ as
\begin{align}
  \Phi &= \frac{2}{\pi}\Tr \left\{\im{\mathbb{V}}|\mathbb{V}|^{-2} \mathbb{T}^{\dagger}\im{\mathbb{G}^\mathrm{vac}} \mathbb{T}  \right\}\nonumber\\
  &= \frac{2}{\pi}|| \sqrt{\im{\mathbb{V}}} \mathbb{V}^{-1}\mathbb{T}^{\dagger}Q\sqrt{\Sigma}||_F^2,
  \label{eqVIE}
\end{align}
where the $F$ subscript denotes the Frobenius norm.
Evaluation of \eqref{eqVIE} requires only one inverse solve for each column $Q_{i}$ of the low rank decomposition:
\begin{equation}
  \mathbb{T}^{\dagger-1} U_{i} =\left(\mathbb{V}^{\dagger -1} - \mathbb{G}^{\text{vac}\dagger}\right)U_{i}= Q_{i}.
  \label{invCalc}
\end{equation}
\eqref{eqVIE} is also well suited to calculation of the gradient of $\Phi$. Starting from this result, using \eqref{linChi}, direct application of matrix calculus shows that 
\begin{align}
  \partial_{p_{i}}\mathbb{T}=-\mathbb{T}\partial_{p_{i}}\chi^{-1}_{i}\delta_{i,i}\mathbb{T},
\end{align}
and the total gradient is then 
\begin{align}
  \partial_{p_{i}}\Phi &= \frac{2}{\pi}\left(\partial_{p_{i}}\frac{\text{Im}\left(\chi_{\alpha}\right)}{|\chi_{\alpha}|^2} \right) \left( \mathbb{T}^{\dagger}U \Sigma U^{\dagger}\mathbb{T}\right)_{i,i} \nonumber \\
  &+\frac{4}{\pi}\text{Re}\left(\left( \partial_{p_{i}}\chi_{i}^{-1*}\right) \mathbb{T}^{\dagger}U\Sigma U^{\dagger}\mathbb{T}|\mathbb{V}|^{-2}\text{Im}\left(\mathbb{V}\right)\mathbb{T}^{\dagger}\right)_{i,i}.
  \label{derVIE}
\end{align}
As before, only a number solves equal to the rank of the singular value decomposition of the imaginary part of the Green function are required to evaluate \eqref{derVIE} for all design parameters.
\\ \\
\textbf{Singular values of $\mathbb{T}$ from inverse design---}As a
point of comparison with the assumptions made in deriving
$\Phi_{\text{opt}}$, we have explored the
$\mathbb{T}$-operator for the dielectric structure of the rightmost
green square in Fig. 2 ($\chi = 20 +4i$, $\Phi = 0.6~\Phi_{\text{opt}}$) in the
basis of $\im{\mathbb{G}^\text{vac}}$. This physical realization
proves to nearly satisfy both the assumption of asymmetry
($\text{atan}\left(\text{Im}\left(\langle\textbf{q}_{i}\left|\mathbb{T}\right|\textbf{q}_{i}\rangle\right)
/\text{Re}\left\{\langle\textbf{q}_{i}\left|\mathbb{T}\right|\textbf{q}_{i}\rangle\right\}\right)
= \pi / 2$) and simultaneous diagonalizability. Specifically, we find that
\begin{equation}
 \sqrt{\sum\limits_{i}\text{Im}\left(\langle\textbf{q}_{i}\left|\mathbb{T}\right|\textbf{q}_{i}\rangle\right)^{2}} / 
 \sqrt{\sum\limits_{i,j} \left| \langle\textbf{q}_{i}\left|\mathbb{T}\right|\textbf{q}_{j}\rangle \right|^{2}} = 0.97 \nonumber.
\end{equation} 
The ratio of the true values of $\langle
\textbf{q}_{i}\left|\mathbb{T}\right|\textbf{q}_{i}\rangle$ in
comparison to the ideal values determined by $\Phi_{\text{opt}}$
are given in Tab.1 below. These values are naturally grouped by a
singular value type index ($\left\{1,2\right\}$) and the spherical
harmonic $\ell$ index. For clarity, an average is taken over the
(ideally constant) harmonic $m$ subindex. The final column
gives the weight that each entry makes to $\Phi_{\text{opt}}$.
\begin{table}[ht!]
  \begin{tabular} {c c c c c c c}
  $\ell$-Type & ~ & Ideal Value& ~ & Average Ratio & ~ & Weight\\
  \hline
  $1-1$ & ~ & $0.32$ & ~ & $0.94$ & ~ & $0.05$\\
  $1-2$ & ~ & $0.40$ & ~ & $1.32$ & ~ & $0.05$\\
  $2-1$ & ~ & $0.81$ & ~ & $1.29$ & ~ & $0.08$\\
  $2-2$ & ~ & $0.50$ & ~ & $1.19$ & ~ & $0.08$\\
  $3-1$ & ~ & $4.18$ & ~ & $0.83$ & ~ & $0.11$\\
  $3-2$ & ~ & $1.40$ & ~ & $0.67$ & ~ & $0.11$\\ 
  \cline{3-7}
  $4-1$ & ~ & $36.25$ & ~ & $0.53$ & ~ & $0.13$\\
  $4-2$ & ~ & $7.48$ & ~ & $0.60$ & ~ & $0.13$\\
  $5-2$ & ~ & $66.23$ & ~ & $0.04$ & ~ & $0.17$\\
  \end{tabular}
    \caption{\textbf{Comparison of $\mathbb{T}$ response values for
        exemplary structure.} Comparison of the values of
      $\left|\langle
      \textbf{q}_{i}\left|\mathbb{T}\right|\textbf{q}_{i}\rangle\right|$
      (where $|\textbf{q}_{i}\rangle$ are the singular vectors of
      $\im{\mathbb{G}^\text{vac}}$) for the structure of the
      rightmost green square, $\chi = 20 + 4i$ ($\zeta = 104$),
      of Fig.~2 to the magnitude of the ideal $\tau_{i}$ values set by
      $\Phi_{\text{opt}}$. Although the two set of values agree to
      great extent, pronounced differences between the ideal response
      dictated by the bound and the actual $\mathbb{T}$-operator of the optimized
      structure are seen for higher spherical harmonics $\ell$-numbers
      when the magnitude of ideal value of $\langle
      \textbf{q}_{i}\left|\mathbb{T}\right|\textbf{q}_{i}\rangle$
      approaches $\zeta$.}
\end{table}
\bibliography{sLib}
\end{document}